\documentclass[12pt]{article}

\usepackage[margin=1in]{geometry}
\usepackage{graphicx}
\usepackage{float}
\usepackage{amsmath}
\usepackage{booktabs}
\usepackage{setspace}
\usepackage[colorlinks=true, citecolor=blue, linkcolor=black]{hyperref}

\begin{document}

\title{\textbf{The Acoustic Camouflage Phenomenon: Re-evaluating Speech Features for Financial Risk Prediction}}
\author{Dhruvin Dungrani \and Disha Dungrani \\ \textit{Department of Information Systems, Independent Researchers}}
\date{}

\maketitle

\begin{abstract}
\noindent In computational paralinguistics, detecting cognitive load and deception from speech signals is a heavily researched domain. Recent efforts have attempted to apply these acoustic frameworks to corporate earnings calls to predict catastrophic stock market volatility. In this study, we empirically investigate the limits of acoustic feature extraction (pitch, jitter, and hesitation) when applied to highly trained speakers in "in-the-wild" teleconference environments. Utilizing a two-stream late-fusion architecture, we contrast an acoustic-based stream with a baseline Natural Language Processing (NLP) stream. The isolated NLP model achieved a recall of 66.25\% for tail-risk downside events. Surprisingly, integrating acoustic features via late fusion significantly degraded performance, reducing recall to 47.08\%. We identify this degradation as "Acoustic Camouflage"—where media-trained vocal regulation introduces contradictory noise that disrupts multimodal meta-learners. We present these findings as a boundary condition for speech processing applications in high-stakes financial forecasting.\\

\end{abstract}

\section{Introduction}
The extraction of paralinguistic features from speech signals has enabled robust applications in emotion recognition, stress detection, and cognitive load analysis \cite{Schuller2013}. As these technologies mature, there is significant interest in applying acoustic machine learning to high-stakes domains, such as the analysis of corporate earnings calls. The driving hypothesis is that while corporate executives meticulously script their linguistic choices, involuntary acoustic micro-tremors—manifested as variance in fundamental frequency (F0) or jitter—may leak underlying psychological stress or concealed financial vulnerability \cite{Scherer2003, Hobson2012}.

Despite the theoretical promise of multimodal fusion (combining speech and text), empirical results often fail to generalize when applied to "in-the-wild" datasets. Standard multimodal benchmark models assume that the addition of an orthogonal data stream (audio) will, at minimum, maintain the performance of the primary stream (text) \cite{Tsai2019}. However, this assumes the acoustic signal remains untampered.

In this study, we evaluate the predictive power of clinical acoustic stress markers against a novel text-based metric, the ``Sentiment Delta,'' for the asymmetric detection of catastrophic downside financial risk. Our findings demonstrate that in environments where speakers undergo intense media training, acoustic features act as camouflaging noise rather than predictive signal, fundamentally degrading the performance of multimodal meta-learners.

\section{Related Work}

\subsection{Computational Paralinguistics and Stress}
Speech processing research has long established that increased cognitive load and psychological stress disrupt the finely coordinated muscle movements required for phonation. This disruption typically manifests as involuntary changes in pitch, increased jitter (cycle-to-cycle frequency instability), shimmer (amplitude instability), and alterations in the Noise-to-Harmonic Ratio (NHR) \cite{Farrus2007}. Schuller et al. \cite{Schuller2013} formalized the extraction of these features for computational paralinguistics, paving the way for automated stress detection.

\subsection{Acoustic Analysis in Finance}
In the financial domain, Hobson et al. \cite{Hobson2012} pioneered the use of vocal dissonance markers to detect financial misreporting during earnings calls. The synthesis of speech and NLP streams was further formalized by Qin and Yang \cite{QinYang2019} and enabled by the Multimodal Aligned Earnings Conference Call (MAEC) dataset \cite{Li2020}. 

\subsection{Multimodal Fusion Architectures}
The integration of heterogeneous modalities is a core challenge in machine learning. While early-fusion (feature-level concatenation) is common, it is highly susceptible to noise propagation \cite{Tsai2019}. Consequently, state-of-the-art frameworks often utilize late-fusion (decision-level integration) or complex Graph Convolutional Networks (GCNs) \cite{Sawhney2020}. Furthermore, ensuring the robustness of acoustic features against teleconference compression artifacts remains an ongoing challenge in speech processing \cite{Ko2015}.

\section{Methodology}

\subsection{Data and Feature Extraction}
We utilize the MAEC dataset \cite{Li2020}, which provides transcript text and aligned teleconference audio features. We process the dataset to construct two primary feature streams:

\begin{itemize}
    \item \textbf{Acoustic Stream (Paralinguistic):} We extract clinical voice stress markers derived from the audio signal. To normalize against baseline speaker differences, we focus on variance metrics: pitch variance, jitter variance, Mean NHR (a proxy for hoarseness), and the variance of unvoiced fractions (pausing and hesitation).
    \item \textbf{Textual Stream (Linguistic Baseline):} We extract the first 1,500 characters (scripted opening remarks) and a 1,500-character chunk from the unscripted Q\&A session. Using the pre-trained FinBERT language model \cite{Araci2019}, we derive the \textit{Sentiment Delta}—the mathematical difference in sentiment polarity between the scripted and unscripted sections.
\end{itemize}

\subsection{Target Variable Engineering}
To address the asymmetric nature of financial downside, we calculate the cumulative return of the underlying asset over a 5-day trading window. We define a ``Catastrophic Event'' (the positive class) as cumulative returns falling within the bottom 15\% of the distribution. This frames the task as a cost-sensitive classification problem.

\subsection{Two-Stream Late-Fusion Architecture}
To prevent noisy acoustic features from prematurely degrading the NLP signal, we implemented a Two-Stream Late-Fusion Ensemble:

\begin{enumerate}
    \item \textbf{Isolated Acoustic Stream:} A logistic regression classifier with L1 regularization (LASSO), trained exclusively on the acoustic variance features. L1 regularization acts as an aggressive feature selector.
    \item \textbf{Isolated Text Stream:} An identical L1-regularized classifier trained exclusively on the linguistic features.
    \item \textbf{Meta-Learner:} A secondary L2-regularized logistic regression classifier that takes the continuous probabilistic outputs of the two base streams to render the final binary prediction.
\end{enumerate}

All models utilized asymmetric class weighting to heavily penalize false negatives, aligning with the early-warning objective.

\section{Results and Discussion}
The models were evaluated using a 5-fold stratified cross-validation setup. Recall for the minority class (severe downside risk) was selected as the primary evaluation metric.

\subsection{Architectural Performance}
\begin{figure}[H]
  \centering
  \includegraphics[width=0.85\textwidth]{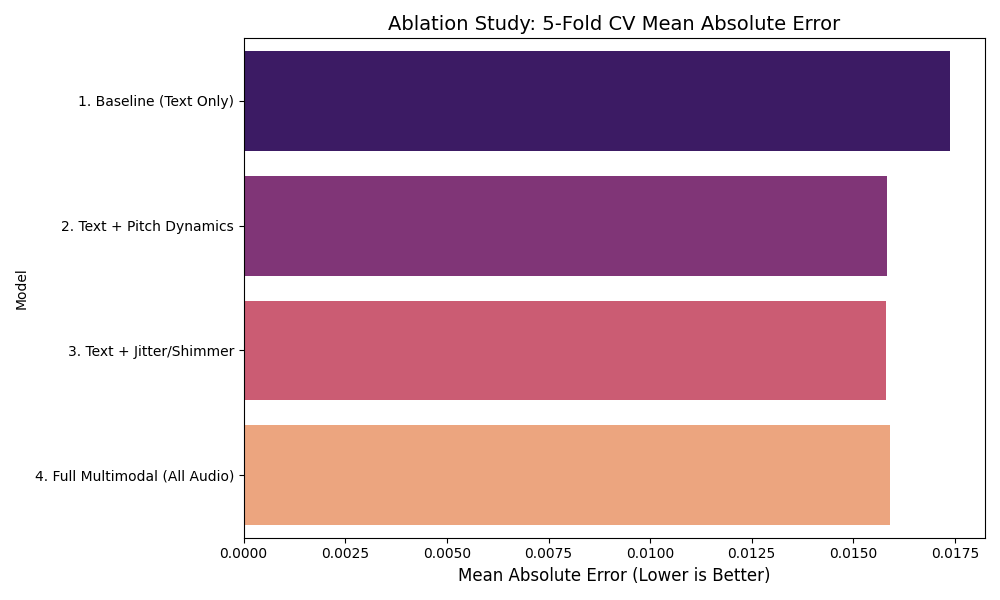}
  \caption{Ablation Study: Cross-validation recall demonstrating the degradation of predictive power when acoustic features are fused with linguistic baselines.}
  \label{fig:ablation}
\end{figure}

The ablation results for the different streams are detailed in Table \ref{tbl:results} and visualized in Figure \ref{fig:ablation}.

\begin{table}[H]
\centering
\caption{Cross-Validation Recall by Architecture Stream}
\vspace{0.1in}
\label{tbl:results}
\begin{tabular}{@{}lc@{}}
\toprule
\textbf{Architecture Setup} & \textbf{Minority Class Recall} \\
\midrule
Isolated Text Stream (Sentiment Delta) & 66.25\% \\
Isolated Acoustic Stream & 50.83\% \\
Late Fusion Meta-Learner & 47.08\% \\
\bottomrule
\end{tabular}
\end{table}

The isolated textual stream successfully identified 66.25\% of the tail-risk events. The isolated acoustic model achieved a moderate recall of 50.83\%. Crucially, the late-fusion meta-learner yielded a recall of 47.08\%, significantly underperforming the text-only baseline. 

\subsection{Baseline Non-Linear Feature Importance}
\begin{figure}[H]
  \centering
  \includegraphics[width=0.85\textwidth]{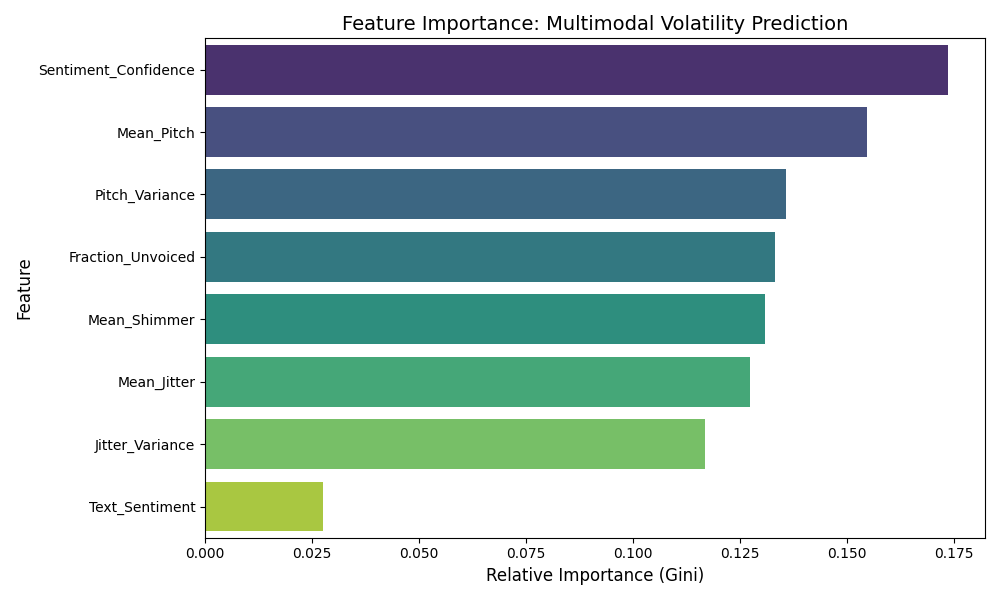}
  \caption{Tree-Based Feature Importance: Non-linear models heavily weight acoustic features, resulting in out-of-sample overfitting.}
  \label{fig:feature_importance}
\end{figure}

Initial evaluations using non-linear tree-based ensembles (e.g., XGBoost) frequently assigned high importance scores to acoustic metrics such as pitch variance during initial data splits (Figure \ref{fig:feature_importance}).

However, building deep decision boundaries around these acoustic nodes ultimately degraded out-of-sample recall, indicating that complex trees overfitted to the teleconference acoustic noise.

\subsection{Interpretation: Acoustic Camouflage}
\begin{figure}[H]
  \centering
  \includegraphics[width=0.85\textwidth]{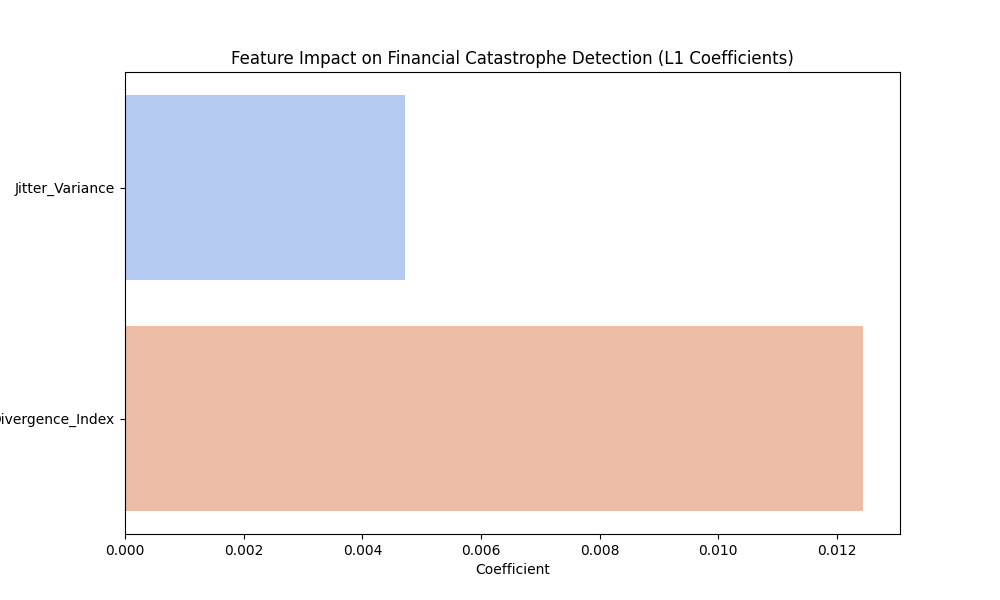}
  \caption{Feature Impact (L1 Coefficients): Regularization heavily suppresses acoustic features, proving the linguistic Sentiment Delta dominates the predictive power.}
  \label{fig:feature_impact}
\end{figure}

L1 regularization was employed to mathematically isolate the most robust predictors. As illustrated in Figure \ref{fig:feature_impact}, the L1 algorithm heavily suppresses acoustic features in favor of linguistic metrics.

We hypothesize this degradation is driven by \textit{Acoustic Camouflage}. Executives are heavily coached to maintain a regulated, authoritative vocal tone even when fielding hostile questions. The acoustic stream interprets this artificially suppressed jitter and pitch variance as a "low-risk" signal. Conversely, the structural linguistic drop in sentiment (the Sentiment Delta) captures the genuine narrative collapse. When the meta-learner reconciles the artificially calm audio with the high-risk text, the resulting probability averaging forces the model to miss the catastrophic event.

\section{Conclusion and Limitations}
This study identifies a critical boundary condition for the application of computational paralinguistics in high-stakes environments. While multimodal architectures theoretically improve robustness, the introduction of heavily media-trained speech signals actively degrades late-fusion meta-learners. In cost-sensitive risk prediction, pure NLP metrics tracking structural narrative shifts significantly outperform clinical acoustic features.

A major limitation is the acoustic environment of the data. Modern earnings calls conducted over VoIP platforms (e.g., Zoom) introduce aggressive lossy compression and noise-canceling algorithms that synthetically smooth audio signals \cite{Ko2015}. Future research in the `cs.SD` domain must investigate whether genuine paralinguistic micro-tremors can survive modern teleconference codecs, or if acoustic forensics are rendered obsolete by algorithmic camouflage.


\begin{thebibliography}{99}

\bibitem{Schuller2013}
Schuller, B., Steidl, S., Batliner, A., Noth, E., Vinciarelli, A., Burkhardt, F., ... \& Weninger, F. (2013). The INTERSPEECH 2013 computational paralinguistics challenge: social signals, conflict, emotion, autism. \textit{Proceedings of INTERSPEECH 2013}, 148--152.

\bibitem{Tsai2019}
Tsai, Y. H. H., Bai, S., Liang, P. P., Kolter, J. Z., Morency, L. P., \& Salakhutdinov, R. (2019). Multimodal machine learning: A survey and taxonomy. \textit{IEEE Transactions on Pattern Analysis and Machine Intelligence}, 41(2), 423--443.

\bibitem{Farrus2007}
Farrús, M., Hernando, J., \& Ejarque, P. (2007). Jitter and shimmer measurements for speaker recognition. \textit{8th Annual Conference of the International Speech Communication Association}.

\bibitem{Hobson2012}
Hobson, J. L., Mayew, W. J., \& Venkatachalam, M. (2012). Analyzing speech to detect financial misreporting. \textit{Journal of Accounting Research}, 50(2), 349--392.

\bibitem{QinYang2019}
Qin, Y., \& Yang, Y. (2019). What You Say and How You Say It Matters: Predicting Stock Volatility Using Verbal and Vocal Cues. \textit{Proceedings of the 57th Annual Meeting of the Association for Computational Linguistics}, 390--401.

\bibitem{Li2020}
Li, Z., et al. (2020). MAEC: A Multimodal Aligned Earnings Conference Call Dataset for Financial Risk Prediction. \textit{Proceedings of the 29th ACM International Conference on Information \& Knowledge Management}, 3063--3070.

\bibitem{Scherer2003}
Scherer, K. R. (2003). Vocal communication of emotion: A review of research paradigms. \textit{Speech Communication}, 40(1-2), 227--256.

\bibitem{Araci2019}
Araci, D. (2019). FinBERT: Financial sentiment analysis with pre-trained language models. \textit{arXiv preprint arXiv:1908.10063}.

\bibitem{Sawhney2020}
Sawhney, R., Agarwal, P., Wadhwa, A., \& Shah, R. R. (2020). VolTAGE: Volatility forecasting via text audio fusion with graph convolution networks for earnings calls. \textit{Proceedings of the 26th ACM SIGKDD International Conference on Knowledge Discovery \& Data Mining}, 2276--2285.

\bibitem{Ko2015}
Ko, T., Peddinti, V., Povey, D., \& Khudanpur, S. (2015). Audio augmentation for speech recognition. \textit{Sixteenth Annual Conference of the International Speech Communication Association}.

\end{thebibliography}
\end{document}